\documentclass[manuscript]{emulateapj}
\usepackage{natbib}
\usepackage{rotating}

\newcommand{\myemail}{e.a.helder@astro-uu.nl}
   \newcommand{\kms}{{km\,s$^{-1}$}}

\slugcomment{Accepted to ApJ May 28, 2011}

\shorttitle{Temperature equilibration behind the shock front}
\shortauthors{Helder et al.}

\begin{document}

\title{Temperature equilibration behind the shock front:\\
 an optical and X-ray study of RCW~86}

\author{E.A. Helder, J. Vink}
\affil{Astronomical Institute, Utrecht University,  P.O. Box 80000, 3508 TA Utrecht, The Netherlands}

\author{C.G. Bassa}
\affil{Jodrell Bank Centre for Astrophysics, School of Physics and Astronomy, The University of Manchester, Oxford Rd, Manchester, M13 9PL UK}
\email{\myemail}

\begin{abstract}
We study the electron-proton temperature equilibration behind several shocks of the RCW~86 supernova remnant. To measure the proton temperature, we use published and new optical spectra, all from different locations on the remnant. For each location, we determine the electron temperature from X-ray spectra, and correct for temperature equilibration between the shock front and the location of the X-ray spectrum. We confirm the result of previous studies that the electron and proton temperatures behind shock fronts are consistent with equilibration for slow shocks and deviate for faster shocks. However, we can not confirm the previously reported trend of $T_{\rm e}/T_{\rm p}\propto 1/v_{\rm s}^2$. 
\end{abstract}


\section{Introduction}
Shocks are ubiquitous in the Universe and occur at various scales: from planetary shocks in our Solar System to large scale shocks in clusters of galaxies. Astrophysical shocks are rather different from shocks on Earth, as typical mean-free-paths for particle-particle interactions in the interstellar medium are of the order of parsecs, which is larger than the scale of the shock transitions of interstellar shocks. The fact that we observe sharp transitions, implies that the existence of the shock must be communicated on scales shorter than this mean-free-path.
From conservation of mass, momentum and energy over the shock front, and assuming an infinite Mach number and no cosmic-ray acceleration, \begin{equation}
kT_{\rm i} = \frac{2(\gamma - 1)}{(\gamma + 1)^2} m_{\rm i} v_{\rm s}^2,
\label{heat}
\end{equation}
in which $v_{\rm s}$ denotes the shock velocity , and $\gamma$ the equation of state of the plasma ($\gamma=5/3$ for a non-relativistic gas). This equation holds for a plasma maximally out of thermal equilibrium, implying an electron temperature  over proton temperature ($T_{\rm e}/T_{\rm p}$) of the ratio of the electron and proton mass ($m_{\rm e}/m_{\rm p} = 1/1836$). For a plasma in thermal equilibrium, equation \ref{heat} reads:
\begin{equation}
kT = <kT_{\rm i}>= \frac{2(\gamma - 1)}{(\gamma + 1)^2} \mu m_{\rm p} v_{\rm s}^2, 
\label{equil}
\end{equation}
in which $\mu$ is the mean particle mass (0.6 for a fully ionized plasma with solar abundances). However, if a shock is accelerating cosmic rays, a cosmic-ray precursor will develop, which heats the incoming medium and lowers the effective Mach number of the shock, in  which case equation \ref{heat} or \ref{equil} are no longer valid \citep[see also,][]{Drury2009,Vink2010}.

In most empirical studies on temperature equilibration behind supernova remnant shocks, $T_{\rm e}$ and $T_{\rm p}$ are determined from Balmer dominated shocks. These shocks are characterized by hydrogen line emission in the Balmer series, leading to a line profile consisting of two superimposed lines. These shocks are non-radiative;  their energy losses in the form of radiation are dynamically negligible \citep[this happens for shock velocities $>$ 200 \kms; e.g. ][]{Draine}. This implies that the hydrogen lines of these shocks are caused by impact excitation, not by recombination. The narrow component of the spectrum is emitted by neutral hydrogen atoms, excited shortly after entering the shock front; its width reflects the temperature of the ambient medium. The broad component is emitted after charge exchange between incoming neutral hydrogen and hot protons behind the shock front. The width reflects the temperature of the protons behind the shock front. Combining the width with the ratio of the flux in the narrow to the flux in the broad component, one can determine the shock velocity and hence the electron temperature \citep[for a recent review, see][]{Heng2010}. This procedure works for shocks which have negligible cosmic-ray acceleration. If a shock is an efficient accelerator, the cosmic-ray precursor ionizes and heats the incoming medium and therewith increases the flux of the narrow component \citep{Raymond2011}. 

The electron temperature for most young supernova remnants can also be measured from the X-ray spectrum, as the thermal component of the X-ray spectrum is shaped by the electron temperature. Electron temperatures measured from X-ray spectra are generally not the electron temperatures right behind the shock front, but rather somewhat further to the inside of the shock, where temperatures have already (partially) equilibrated through Coulomb interactions. The amount of equilibration is a function of $n_{\rm e}t$: the ionization age, that can be determined from the X-ray spectrum simultaneously with the electron temperature, from the ratio of the fluxes in different spectral lines from the same species. 

\cite{Rakowski2003} studied shocks of the supernova remnant DEM L71 at both optical and X-ray wavelengths. The remnant has shock velocities between 555 and 980 \kms. Correcting for temperature equilibration effects, \cite{Rakowski2003} showed that $T_{\rm e}/T_{\rm p}$ decreases with increasing shock velocity. 

From a theoretical point of view, it is not well determined how much electron-proton temperature equilibration would to expect behind a shock front. Empirical studies of shocks with negligible cosmic-ray acceleration, show that electron and proton temperatures are equilibrated for slow shocks ($v_{\rm s}<$ 400 \kms), whereas the degree of thermal equilibration decreases for faster shocks \citep{Rakowski2003,Ghavamian2007,Adelsberg}. More specifically, a relation of  $T_{\rm e}/T_{\rm p}\propto v_{\rm s}^{-2}$ has been suggested for shocks with $v_s > 400$ \kms\  \citep{Ghavamian2007}. This directly implies that $T_{\rm e}$ is nearly constant closely behind the shock front  \citep[i.e. apart from temperature equilibration effects, equation 2 in][]{Ghavamian2002}, independent of the shock velocity (equation \ref{heat}). The origin of this relation is attributed by \citet{Ghavamian2007} to so-called lower hybrid waves ahead of the shock, caused by a (moderate) cosmic-ray precursor. Note that \cite{Adelsberg} could not reproduce $T_{\rm e}/T_{\rm p}\propto v_{\rm s}^{-2}$, using published data of H$\alpha$-spectra as well, but with different models for interpreting H$\alpha$ spectra.  
\citet{Bykov1999} and \cite{Bykov2004} point out that for Mach numbers $M <  \sqrt{m_{\rm p}/m_{\rm e}} = 43$  the electrons ahead of the shock have a typical thermal velocity
exceeding the shock velocity, which may lead, through non-resonant interactions with large-amplitude turbulent fluctuations in the shock transition region, to collisionless
electron-heating and acceleration. Interestingly, assuming a typical sound speed of the ambient medium of 10~km/s, $M=43$ corresponds to a shock speed of $\sim 400$~km/s.
For a cosmic-ray accelerating shock it is even more complex, as the incoming material is already preheated by a cosmic-ray precursor, leading to a lower Mach number at the main shock. This could mean that for cosmic-ray dominated shocks the electron-proton temperatures could be equilibrated again.
 
Here, we investigate the electron-proton temperature equilibration behind several shock fronts of RCW~86, using X-ray (XMM-Newton/MOS) and optical data. The shocks of the supernova remnant RCW~86 are particularly well suited for this study, as the ionization age of the plasma in parts of the remnant is low: $\sim5\times10^9~{\rm cm}^{-3}{\rm s}$ \citep[e.g. ][]{Vink1997}, this value is low compared to the ionization age of most other remnants \footnote{A plasma can be regarded as being in collisional equilibration for $n_{\rm e}t$ of $\gtrsim 10^{12}~{\rm cm}^{-3}~{\rm s}$. }. The low value for $n_{\rm e}t$ makes the measured electron temperature a more direct indication for the electron temperature at the shock front.  

Additionally, the shocks of RCW~86 have a large variety in speeds, as the progenitor exploded in its own wind-blown bubble \citep[e.g. ][]{Vink1997} and parts of the remnant have already hit the dense rim of this bubble. As a result, the southwest (SW) rim has slowed down; its X-ray emission is mostly thermal and the shock velocity is $\sim$ 500~\kms\ \citep{Ghavamian2001}. The X-ray emission of the northeast (NE) shock is dominated by X-ray synchrotron emission \citep{Vink2006}, and its shock velocity is measured to be 6000 $\pm$ 2800 \kms\ \citep{Helder2009}. However, its proton temperature is low, which is an indication for efficient cosmic-ray acceleration.

Furthermore, the mostly sub-solar abundances of the thermal X-ray emission at the shock front of RCW~86 indicate emission from shocked ambient medium \citep[rather than shocked ejecta, e.g., ][]{Vink1997, Borkowski2001, Rho2002}, tying the fitted parameters of the X-ray spectrum to the physics at the shock front.

\section{Data and results}
\subsection{VLT/FORS2}\label{FORS}

In this study, we use both published and new optical observations. The observation taken from the SW of the remnant was previously published in \cite{Ghavamian2001} and the observation from the NE rim was published in \cite{Helder2009}. The parameters of the north (N) and east (E) observations are taken from \cite{Ghavamian2007} and the northwest (NW) location from \cite{GhavamianPhD}. The exact locations of the N and E spectra have been kindly provided to us by Parviz Ghavamian (private communication). We also use parameters of a spectrum published in \cite{Long}, this spectrum is taken from the northern region as well. Table 2 summarizes the line widths used in the present paper. 

\begin{figure*}[!t]
\begin{center}
\includegraphics[angle = 0, width=0.7\textwidth]{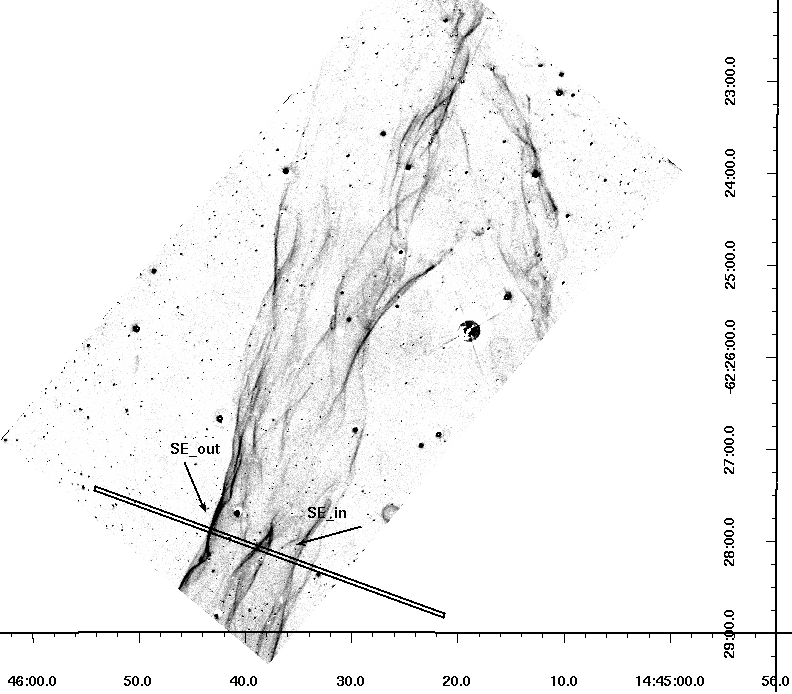}  \\
\caption{Southwest H$\alpha$ filament of RCW~86, as observed with VLT/FORS2.  Overlaid is the slit position used for the spectra in Fig. \ref{VLT_spectra}.} 
             \label{SW_VLT}%
\end{center}
\end{figure*}

The optical image of Fig. \ref{SW_VLT} has been obtained with VLT/FORS2 as pre-image for the spectral observation, both in 2007 (programme  ID 079.D-0735). The image is a combination of three exposures of 200s, through a narrowband H$\alpha$ filter (\texttt{H\_Alpha+83}). To correct for stellar light, three similar images through a narrowband H$\alpha$ filter, shifted 4500 \kms\ (\texttt{H\_Alpha/4500+61}) have been taken.

\begin{figure*}[]
\begin{center}
\begin{tabular}{c c c}
\includegraphics[angle = 0, width=0.45\textwidth]{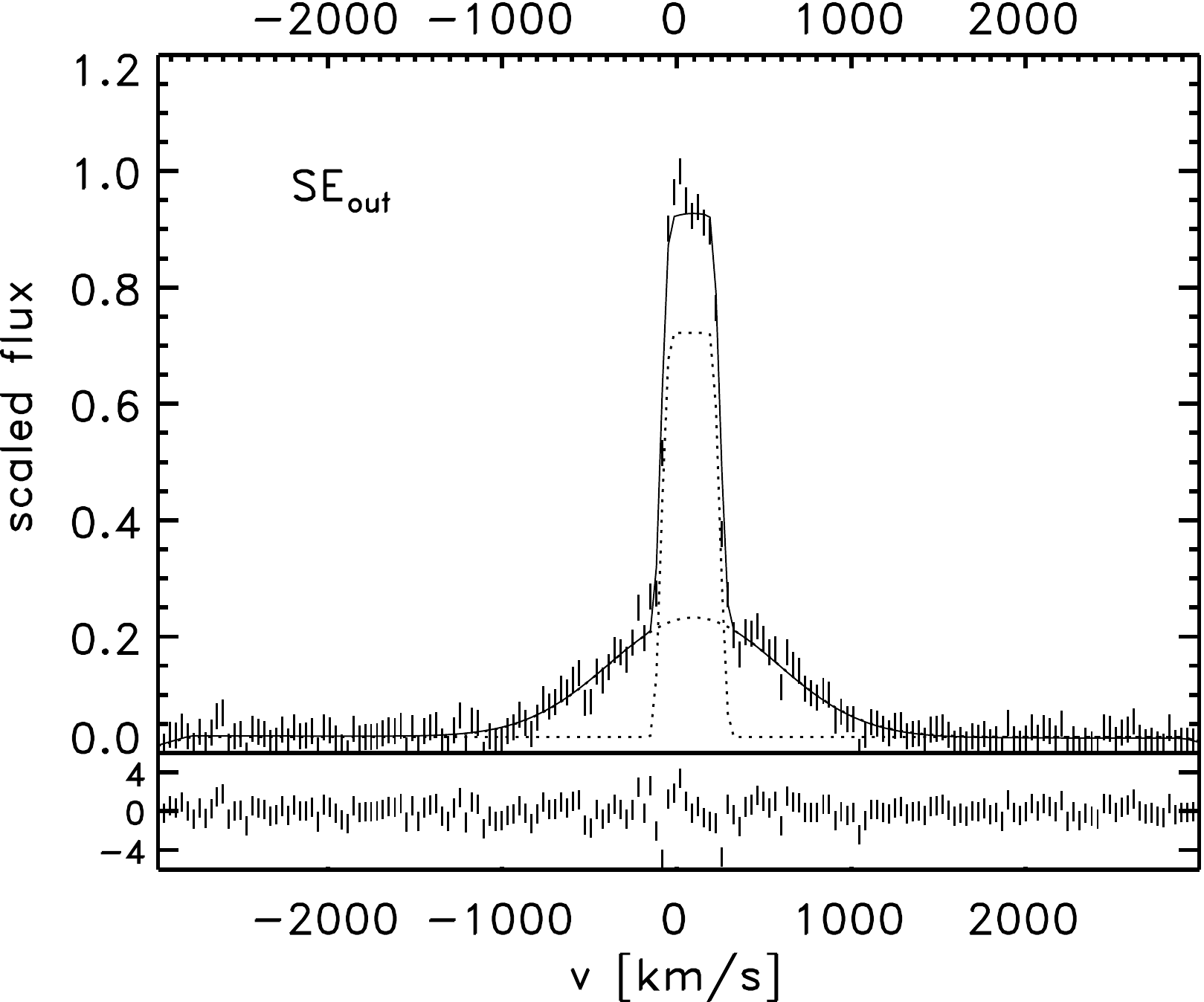} &
\includegraphics[angle = 0, width=0.45\textwidth]{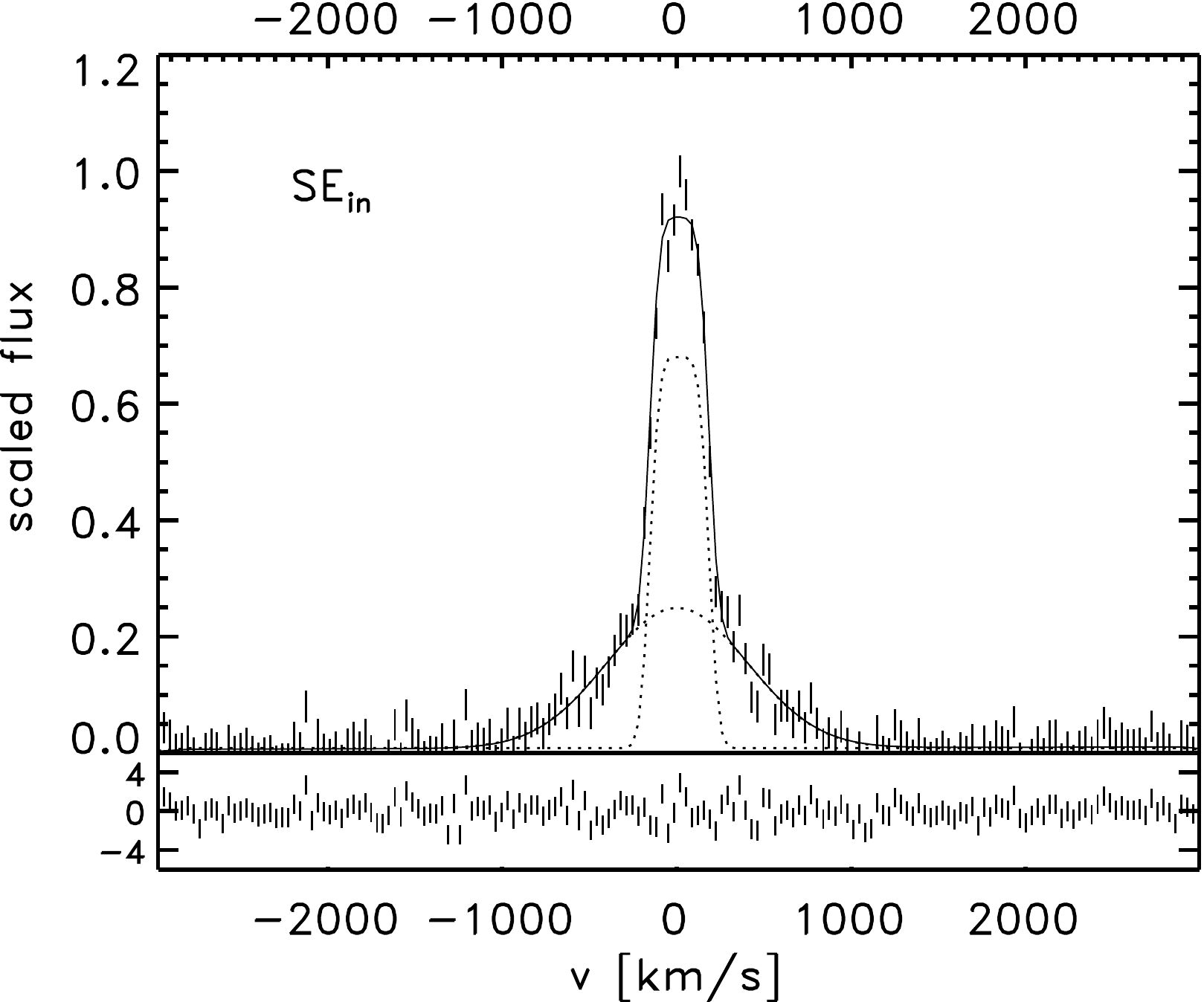} \\
\end{tabular}
\caption{ H$\alpha$ line in the spectrum of the SE$_{\rm out}$ region (left) 
and for the SE$_{\rm out}$ region (right). Overplotted in both figures are the the individual broad and narrow components (dotted lines) and the total fit (solid line).}
	\vskip 7mm
              \label{VLT_spectra}%
\end{center}
\end{figure*}

Figure \ref{SW_VLT} shows the location of the slit, used to obtain the spectra.  The spectra of the southeast (SE) part of the remnant were taken in a single 2733 s exposure, using a 1200R grism and a 2.5\arcsec\ slit. This setup resulted in a spectral resolution of 350 \kms, which is sufficient for resolving the broad component, however, the narrow component of the line probably remains unresolved. The data were reduced with standard data reduction steps, as described in \cite{Helder2009}, resulting in the spectra shown in Fig. \ref{VLT_spectra}.  

The slit crossed several filaments (Fig. \ref{SW_VLT}), and we here present spectra of the outer and inner filaments. We focus on the line width of the broad component, as this is directly related to the post-shock proton temperature. We fit the spectra with two Gaussian lines, convolved with the resolution of the instrument set-up. Table \ref{summ} lists the line widths obtained from the eastern (SE$_{\rm out}$) and western (SE$_{\rm in}$) spectrum in the slit (Fig. \ref{SW_VLT}). As the signal-to-noise of the spectrum of the innermost filament is rather low, we used both the inner and middle filament for the SE$_{\rm in}$ spectrum. A fit to the innermost filament solely revealed a width of the broad line similar to the width found for the combined spectrum. 

\begin{center}
\begin{minipage}[h!]{0.9\textwidth}
\smallskip
\begin{tabular}{l|c|c|c}\noalign{\smallskip}
Location &   Broad H$\alpha$ width& proton T & electron T\footnote{At the shock front, derived in this study}  \\
 & [\kms] & [keV] & [keV]\\
\hline
NE & 1100$\pm$ 60\footnote{\cite{Helder2009}}  & $2.3\pm0.3^a$ &  $>$1.5  \\ 
SE$_{\rm in}$ & $920\pm50^d$ & $1.6\pm0.2^b$ & $0.75\pm0.15$\\
SE$_{\rm out}$ & $1120\pm40^d$ & $2.4\pm0.2^b$ &   $<0.05$ \\ 
SW & $562\pm18$\footnote{\cite{Ghavamian2001}} & $0.60\pm0.04$\footnote{Derived in this study} & $<0.05$\\
N & 325$\pm$10 \footnote{\cite{Ghavamian2007}} & 0.2$\pm 0.02$ & 1.95$\pm0.23$\\
N & 680$\pm$70\footnote{\cite{Long}}&$0.9\pm0.2^d$ & 1.0$\pm0.2$\\
E & 640$\pm$ 35 $^e$ & 0.77$\pm$ 0.04 & 0.78$\pm$0.13\\
NW & 580$\pm$ 18\footnote{\cite{GhavamianPhD}} &0.64$\pm$0.04 & 0.27$\pm$0.08\\
\noalign{\smallskip}
\end{tabular}\label{summ}
 \end{minipage}
\end{center}

\subsection{XMM-Newton}\label{dataXMM}

\begin{figure}[!t]
\begin{center}
\includegraphics[angle = 0, width=0.45\textwidth]{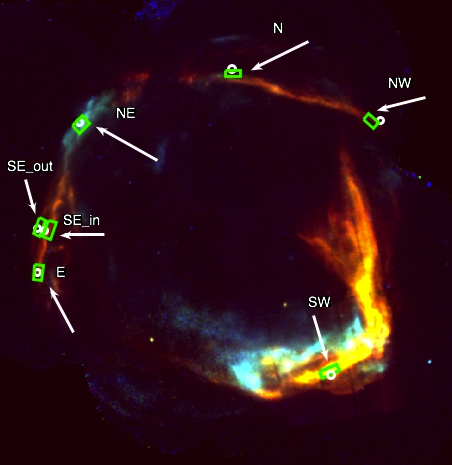}  \\
\caption{RGB image obtained with XMM-Newton. Shown are the regions from which the H$\alpha$ lines have been obtained (white circles). The green boxes show regions from which we extracted the corresponding X-ray spectra. In this image, red indicates 0.5 to 1.0 keV, green 1.0-2.0 keV and blue 2.0-4.0 keV.  } 
	    \label{XMM_regions}%
\end{center}
\end{figure}

\begin{figure}[!t]

\includegraphics[angle = 0, width=0.45\textwidth]{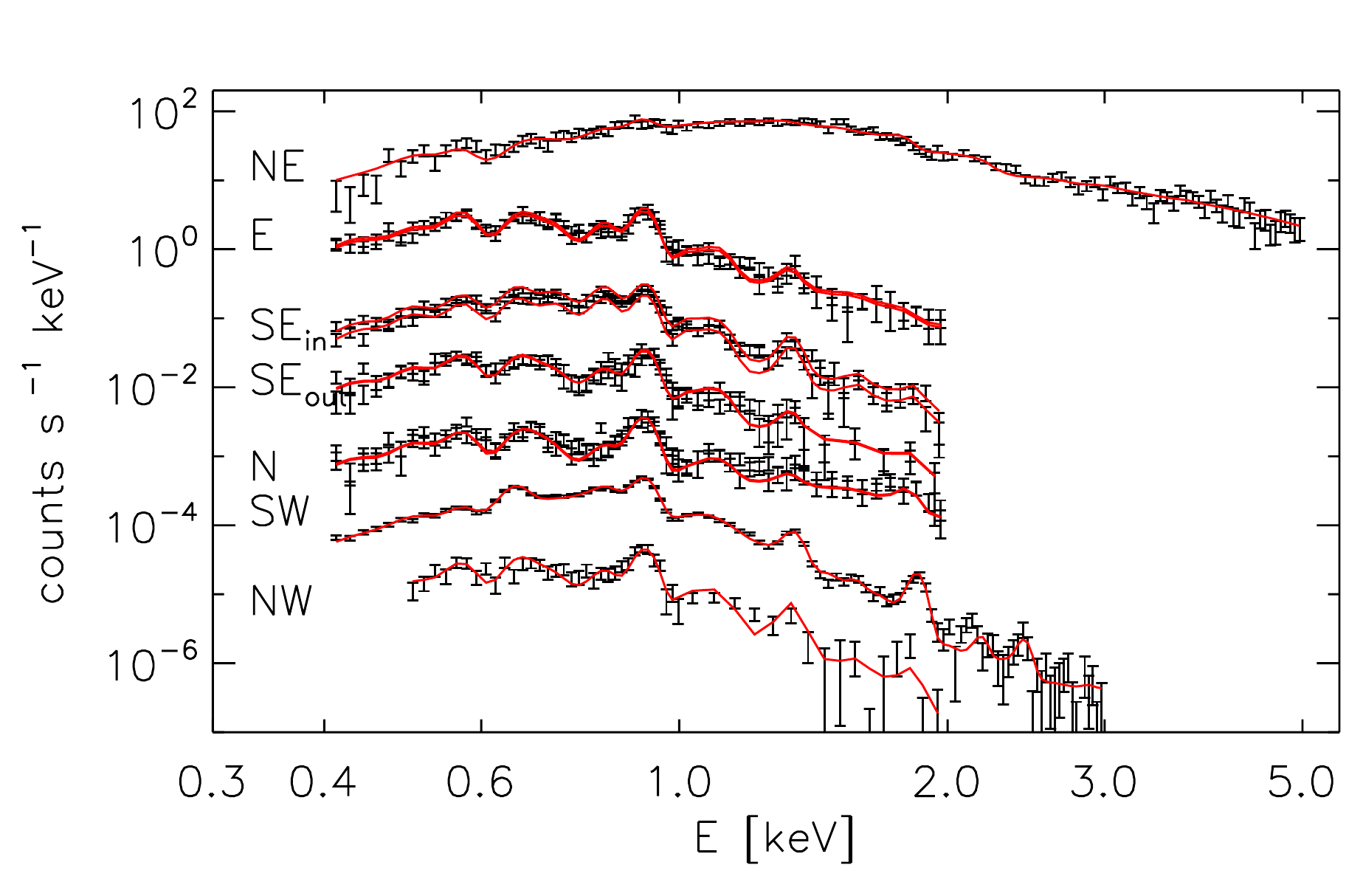}  \\
\caption{Spectra, corresponding to the regions from Fig. \ref{XMM_regions}. Red lines are the best fitting models. The NE spectrum is multiplied with 1000, E with 20, SE$_{\rm in}$ with 1, SE$_{\rm out}$ with 0.3, N with 0.06, SW with 0.0003 and  NW with 0.0007. } 
             \label{Xrayspectra}%
\end{figure}

We determined electron temperatures from X-ray spectra, taken with the XMM-Newton/EPIC MOS instruments \citep{Turner2001}. We chose for the MOS instruments as they have a higher spectral resolution than the XMM-Newton/EPIC pn CCDs \citep{Struder}. The observations for the E, SE$_{\rm in}$ and SE$_{\rm out}$ spectrum were taken on August 13 (ObsId 0504810201, 75 ks), the NE on July 28 (ObsId 0504810101, 117 ks), the N and NW on August 25 2007 (ObsId 0504810301, 74 ks ) and for the SW spectrum on August 23 2007 (ObsId 0504810401, 73 ks). The spectra (shown in Fig. \ref{Xrayspectra}) were extracted at locations close to the regions for which we have H$\alpha$ spectra, as indicated in Fig. \ref{XMM_regions}, using the XMM SAS data reduction software, version 1.52.8. Unfortunately, the position angle of the XMM telescope was chosen such that only the EPIC MOS2 instrument observed the region of interest for the NE, NW and SW rims. For the other spectra, we used both the MOS1 and MOS2 instruments. 

We fitted the extracted spectra with a non-equilibrium ionization (NEI) model \citep{Kaastra1993}, combined with an absorption model, using the SPEX spectral fitting software version 10.0.0 \citep{spex}, using the maximum likelihood statistic for Poisson distributions \cite[C-statistic,][]{Cash1979}. This statistic is more appropriate than the classical $\chi^2$-method for fitting spectra which contain bins with few counts \citep{Wheaton}. For more counts, this statistic asymptotically approaches the $\chi^2$-statistic.
For most of the spectra, this method was relatively straightforward. Some regions needed some special attention, as described below. The NE spectrum needed an additional power-law component, to account for the synchrotron emission present in the spectrum \citep{Vink2006}. Including the power law to account for the synchrotron emission results in an unconstrained electron temperature, with a nominal electron temperature of 37 keV. However, the continuum emission is dominated by synchrotron radiation and the temperature is therefore mostly determined by the very weak line emission. \cite{Vink2006} encountered similar problems but concentrated on a slightly different extraction region, which had more thermal emission. They reported $T_{\rm e} = 6.7 \pm 2.6 $ keV. The high, but unconstrained electron temperature we found for this study, and the high value reported by \cite{Vink2006} suggest that the electron and proton temperature are likely close to equilibration. However, in order to be conservative, we take now the 2$\sigma$ lower limit of \citealt{Vink2006} (1.5 keV) as a lower limit on the electron temperature.  

The SW spectrum can not be fitted adequately with a single NEI component (Cash statistic/d.o.f. = 6.4).  Subsequent analysis with two NEI components with approximately solar abundances , which we coupled for the components, resulted in a Cash statistic/d.o.f. = 4.5. We carried out a different approach carried out by constructing a second NEI component, consisting of pure metals. We constructed this component by increasing the abundances of O, Ne, Mg, Si, S and Fe with a factor of 10$^7$, mimicking a pure metal plasma \citep[a similar procedure was adopted for the 0519-69.0 remnant in the LMC, ][]{Kosenko2010}. We found Cash statistic/d.o.f. = 1.6 for this approach. We found a high T$_{\rm e}$ (2.2 $\pm 0.3$ keV) and $n_{\rm e}$ ($1.4 \pm 0.1 \times 10^{10}$ cm$^{-3}$s) for the pure metal plasma component, and a low T$_{\rm e}$ and $n_{\rm e}$ for the other component (Table \ref{XMMspectra}). This, together with the complex, layered structure of the remnant in the SW, indicates that the pure metal component possibly represents a shocked ejecta component in the spectrum and the low metallicity component is probably tied to the shocked ambient medium. We note that a strong ejecta component with alpha elements was not reported before for this remnant, but several studies suggested the presence of Fe-rich ejecta giving rise to low ionization Fe-K lines \citep{Bamba2000,Borkowski2001,Yamaguchi}. A follow-up studies with a deeper observation is necessary to shed further light on the emission from this region and investigate whether there is indeed an ejecta component present in this region.

The SE$_{\rm out}$ spectrum is fitted with the Fe abundance tied to unity, as otherwise this abundance would rise to unrealistically high values. We note that releasing this component during the fitting procedure will result in an even lower electron temperature (0.3 keV). 

Table \ref{XMMspectra} lists the parameters of the best-fitting spectral model.

\smallskip

\begin{sidewaystable*}

\caption{The best-fit parameters for XMM-Newton MOS spectra. The errors are $1\sigma$. }
\begin{tabular}{l|c|c|c|c|c|c|c}\noalign{\smallskip}
Parameter& NE\footnote{MOS2 data} &   SE$_{\rm in}$\footnote{based on both MOS1 and MOS2 data}&SE$_{\rm out}\,^b$& SW$\,^a$ \footnote{Values quoted are for the low $kT_e$, low abundance component. The second component had EM $= 1.2 \pm 3 \times 10^{-6}$, $kT_{\rm e} = 2.2\pm 0.3$, $n_{\rm e}t = 14\pm 1$,  [O] $= 2.1_{-1.7}^{+13} \times 10^{7}$, [Ne]$= 6\pm3 \times 10^{7}$, [Mg]$=6^{+20}_{-4} \times 10^{6}$, [Si] $= 3.4 ^{+7.4}_{-3.4}\times 10^{6}$, [S] $= 2.5 \pm 1.8 \times 10^6$ and [Fe] $= 4 ^{+16}_{-1.7}\times 10^{6}$. All units are the same as in the table. } & E$\,^b$ & N$\,^b$&NW$\,^a$\\
\hline
EM \footnote{The emission measure (EM) is defined as $\int n_{\rm e}n_{\rm H}dV/d^2$ in units 10$^{55}$ {\rm cm}$^{-3} ({\rm 2.5\ kpc})^{-2}$ }& $0.13\pm0.05$ & $4.4\pm 0.7$& $4 \pm 2  $ & $360\pm 240$&6.0$\pm 0.7 $&4.2$\pm 0.3 $& $20^{+80}_{-9}$\\ 
$kT_e$ (keV)& $37^{+27}_{-19}$ &$0.9 \pm 0.1$ &$0.6 \pm 0.1$& $0.18 \pm 0.01$&$0.8\pm 0.1$&$1.0\pm 0.2$& 0.27$\pm$ 0.08\\
$n_{\rm e}t$ (10$^9$ cm$^{-3}$ s)& $0.24\pm 0.03$&$7.6 \pm 0.7$ &$6\pm 1$& $6.9 \pm 0.5$&$4.2 \pm0.7$&$3.1 \pm0.5$&2.0$^{+4}_{-0.7}$\\
O\footnote{Abundances are relative to solar values \citep{Anders1989}} & 1\footnote{A `1' indicates abundances fixed to solar abundances} &$0.47\pm 0.03$&$0.46\pm 0.05$& $0.51\pm 0.2$&$0.52\pm 0.2$& $0.38\pm 0.05$&1\\
Ne & 1&$0.63\pm 0.05$&$0.52\pm 0.07$&  $4.14\pm 1.6$&$0.54\pm0.02$&$0.44\pm0.06$&1\\
Mg & 1&$0.40\pm 0.05$ &$0.25\pm 0.08$ & $0.7^{+1.4}_{-0.7}$&$0.27^{+0.2}_{-0.05}$&$0.15\pm0.09$ &1\\
Si & 1&$0.3\pm 0.1$&$0.49 \pm 0.4$ & $0^{+12}_{-0}$&$0.05^{+0.2}_{-0.05}$&$0.9\pm0.5$&1\\
S & 1 & $0^{+1.4}_{-0}$ & $0^{+10.4}_{-0}$ & $45^{+1205}_{-45}$&$35.8^{+119}_{-35.8}$& $0^{+50}_{-0}$&1\\
Fe &1 &$0.5\pm 0.1$&$1 $ & $298^{+153}_{-91}$&$2.7\pm0.8$&$2.6^{+5}_{-1.6}$&1\\
PL norm. \footnote{power law normalization [$10^{44}$ ph s$^{-1}$ keV$^{-1}$]}& $7.0\pm 0.3$ &-- &--&--&--&--&--\\
$\Gamma$& $2.9 \pm 0.1$ &-- &--&--&--&--&--\\
$N_{\rm H}$ (10$^{21}$ cm$^{-2}$)& $4.7\pm 0.2$& $2.8\pm 0.2$&$3.6\pm 0.4$&$5.1\pm 0.4$&$3.6\pm 0.2$&$4.2\pm 0.3$& 5.5$\pm$ 1\\
$C$-statistics/dof& 131/117 & 347/136& 197/137 & 145/88&202/137&256/136& 84/52\\

\noalign{\smallskip}
\end{tabular}\label{XMMspectra}
\end{sidewaystable*}

\section{Discussion}
\subsection{XMM-spectra}
The parameters in Table \ref{XMMspectra} are in general consistent with parameters obtained in previous work \citep{Rho2002,Vink2006}. In the NE, the dominant contribution of synchrotron radiation makes it difficult to determine the electron temperature adequately. 
 However, we note that the high contribution of non-thermal X-ray emission to the spectrum makes it difficult to accurately determine the parameters of the thermal component. Our plasma parameters for the SW spectrum differ from previously determined \citep{Rho2002}, this is probably because we fit the spectrum with two components, of which one probably represents the ejecta and one the shocked ambient medium, whereas Rho et al. used a single component. We choose our region right behind a Balmer dominated filament, which is non radiative and hence has a relatively low density. Additionally, the plasma from the ambient medium was probably only recently shocked. Note that this solution is not necessarily unique, but given the high  C-statistic values for the alternative models, we regard this model as an adequate representation of the spectrum.
The low value for the  $n_{e}t$ for the ambient medium component therefore seems appropriate. Alternatively, if we choose our temperature similar to \cite{Rho2002}, $T_{\rm e}$ and $T_{\rm p}$ are close to equilibration. 

Problems with fitting the spectrum around 1.2 keV were also encountered by \cite{Kosenko2008} for the 0509-67.5 supernova remnant. These deviations are possibly caused by uncertainties in the atomic data base of SPEX, presumably by the Fe-L line complex. 
The sub-solar abundances in all fits show that we indeed fit X-ray spectra from shocked ambient medium as opposed to metal-enhanced shocked ejecta. This confirms that the measured electron temperatures are related to the proton temperatures at the shock front.

\subsection{Proton temperatures}

The width of the broad component is primarily a function of the post-shock proton temperature, but modified by the velocity dependent cross sections for charge and impact excitation. This causes the shape of the broad component to deviate from a perfect Gaussian. Nevertheless this remains a decent approximation \citep[see][and references therein]{Adelsberg}.  Fig. 5 of \cite{Adelsberg} shows that up to 2000 km/s, the FWHM of the broad line increases linearly with the shock velocity. This shows that interpreting the line width in terms of proton temperature through $kT_{\rm p} = m_{\rm p}\sigma^2$ is a decent approximation \citep[with $\sigma = {\rm FWHM}/\sqrt{8\ln 2}$ in \kms, ][Table \ref{summ}]{Rybicki,Heng2010}.

Table \ref{summ} shows that $T_{\rm p}$ for SE$_{\rm out}$ is higher than $T_{\rm p}$ for the SE$_{\rm in}$ spectrum, implying a higher shock velocity (equation \ref{heat}). Note that SE$_{\rm out}$ lies outward of SE$_{\rm  in}$, which is suggestive for a higher shock velocity as well. For the Northern region of RCW~86, we have the parameters of two H$\alpha$ spectra \citep{Long, Ghavamian2007} at nearly the same location, which appear to have significantly different proton temperatures (0.2 and 0.9 keV respectively). As we do not know which value best represents this region, we plotted both values for $T_{\rm p}$ with one corresponding $T_{\rm e}$ in Fig. \ref{temperatures}. We note that the measured widths of the broad H$\alpha$-lines tend to vary a lot (from 325 to 543 km/s) in the Northern region \citep{GhavamianPhD}, making it unclear to which $T_{\rm p}$ the electron temperature relates (probably a combination). Also, we note that $T_{\rm p}$ appears to vary significantly along the eastern rim of the remnant  \citep{GhavamianPhD}.

\subsection{Temperature equilibration at the shock front}

\begin{figure}[!t]
\begin{center}
\includegraphics[angle = 0, width=0.45\textwidth]{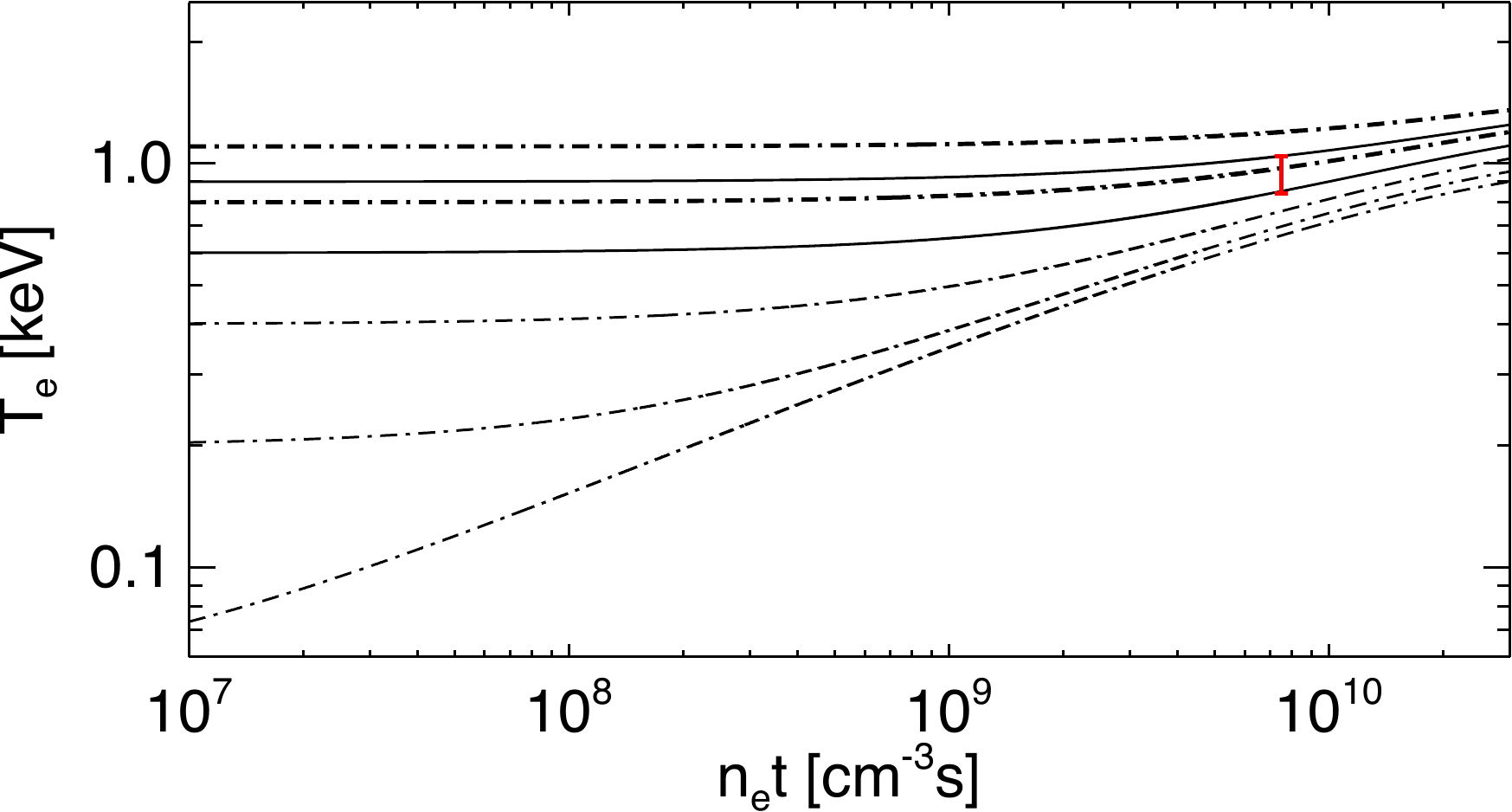}  \\
\caption{Figure with temperature histories for the SE$_{\rm in}$ location ($T_{\rm p}=1.6$~keV), as function $n_{\rm e}t$ for different $T_{\rm e}$-values at the shock front. From top to bottom, the lines indicate a $T_{\rm e}$ at the shock front of 1.10, 0.95, 0.80, 0.60, 0.40, 0.20 and 0.05 keV. } 
             \label{net}%
\end{center}
\end{figure}

The proton temperature is obtained from optical spectra directly behind the shock fronts, as the neutral hydrogen will ionize quickly after entering the shock front. To obtain an X-ray spectrum with sufficient signal-over-noise, we extracted over a larger region of the remnant than the region from which the proton temperature was determined. Also, the spatial resolution of the EPIC-MOS instruments (6\arcsec) prevents us from extracting an X-ray spectrum from very close to the shock front.

\begin{figure}[!t]
\begin{center}
\includegraphics[angle = 0, width=0.45\textwidth]{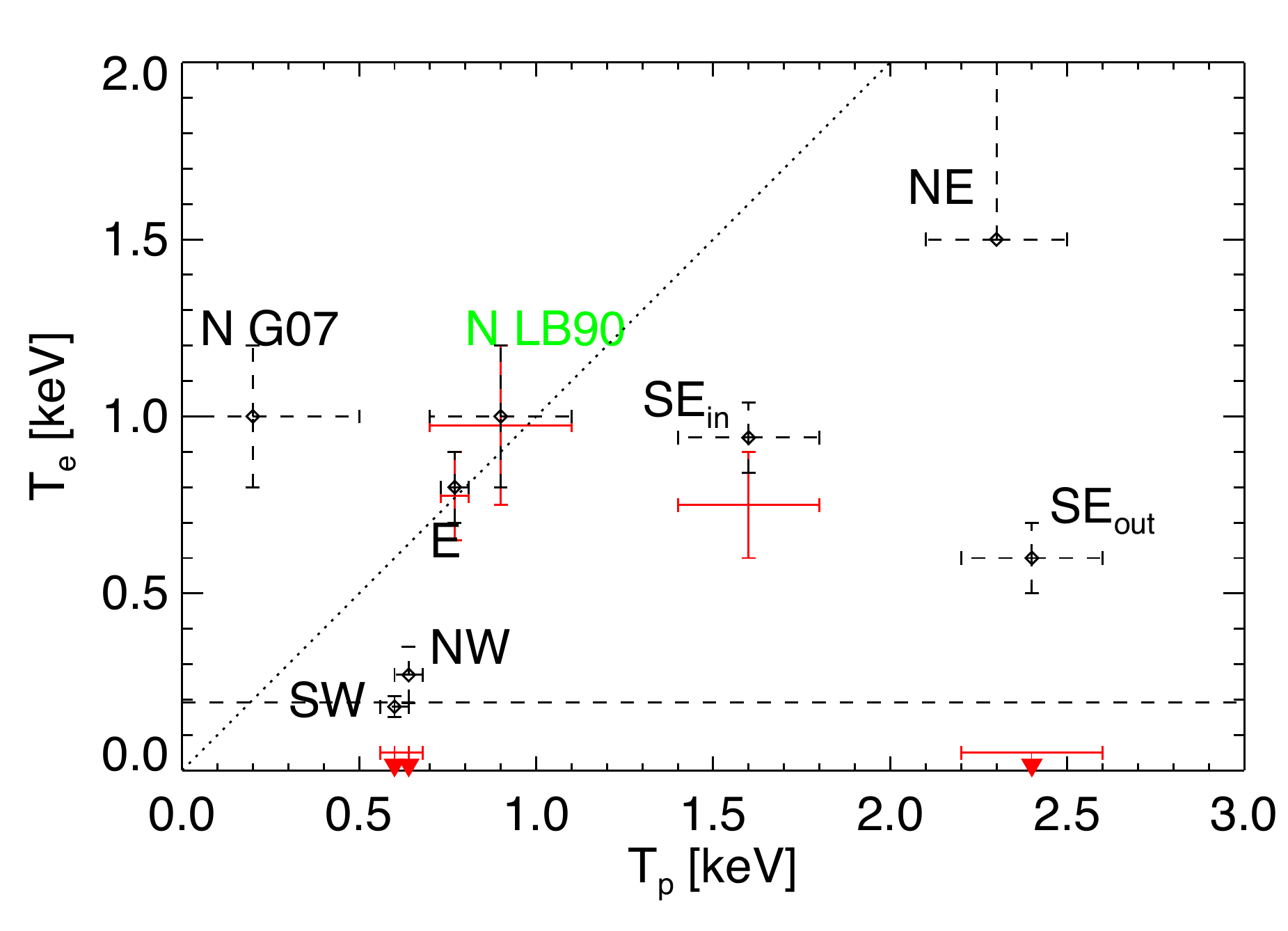}  \\
\caption{Proton and electron temperatures ($x$ and $y$-axis respectively) from Table \ref{summ} and \ref{XMMspectra}. The black data points show the electron temperature as calculated from the X-ray spectra. The red data points indicate electron temperatures at the shock front after correcting for Coulomb equilibration. The dotted line indicates $T_{\rm p} = T_{\rm e}$. The dashed line indicates the electron temperature as calculated for  a 400 \kms\ shock with thermal equilibrium behind the shock front; the relation found by \cite{Ghavamian2007} would be on this line. The points labeled `N G07' and `N LB90' refer to the same region in the North, with $T_{\rm p}$ based on parameters of \cite{Ghavamian2007}  and \cite{Long} respectively.} 
	  \label{temperatures}%
\end{center}
\end{figure}

This means that the electron and proton temperatures will have undergone some equilibration at the extraction region of the X-ray spectrum, implying that the obtained electron temperatures are upper limits (black data points in Fig. \ref{temperatures}).
To compare this to the proton temperatures behind the shock front, we need to determine the amount of heating the electrons experienced. We assume that the electrons are solely heated by Coulomb interactions. \cite{Spitzer1965} derived the coupled differential equation for how fast species i and j with different temperatures ($T_{\rm i}$ and $T_{\rm j}$) equilibrate as function of time ($t$):

\begin{equation}
\frac{\rm{d}T_{\rm i}}{{\rm d}t}=\frac{T_{\rm j} -T_{\rm i}}{t_{\rm eq~(i,j)}},\\
t_{\rm eq~(i,j)} = 5.87\frac{A_{\rm i} A_{\rm j}}{n_{\rm j}Z_{\rm i}^2Z_{\rm j}^2\ln(\Lambda)}\bigg(\frac{T_{\rm i}}{A_{\rm i}}+\frac{T_{\rm j}}{A_{\rm j}}\bigg)^\frac{3}{2}.
\label{coulomb}
\end{equation}
For species with atomic numbers $A_{\rm i}$ and $A_{\rm j}$, charge $Z_{\rm i}$ and $Z_{\rm j}$, number density $n_{\rm j}$ and masses $m_{\rm i}$ and $m_{\rm j}$. $\Lambda$ is given by
$$\Lambda=\frac{3}{2Z_{\rm i}Z_{\rm j}e^3}\bigg(\frac{k^3T_{\rm i}^3}{\pi n_{\rm e}}\bigg)^\frac{1}{2}.$$

Solving the coupled differential equation given by equation \ref{coulomb} for electrons, protons, He, O, Si and Fe (assuming solar abundances), we calculate temperature histories up to the measured $n_{\rm e}t$ for the measured proton temperature, for different electron temperatures at the shock front (Fig. \ref{net}). This enables us to determine the electron temperature at the shock front (Table \ref{summ} and red data points in Fig. \ref{temperatures}). 

Equation \ref{heat} shows that the proton temperature is related to the shock velocity. However, this is based on standard assumptions, assuming only a gaseous component. In contrast, supernova remnants have been suggested to efficiently accelerate particles which will alter the standard Hugoniot relations \citep{Vink2010}. The presence of X-ray synchrotron and TeV radiation indicates that RCW 86 accelerates particles to high energies \citep{Bamba2000, Borkowski2001, Vink2006, AharonianRCW}, whereas the H$\alpha$ line width in the NE shows that in this particular region cosmic-ray acceleration must be efficient, with a post-shock pressure contribution of more than 50\% \citep{Helder2009}. 

When a shock accelerates cosmic rays, the cosmic rays create a cosmic-ray precursor ahead of the shock, pushing out and pre-heating the incoming medium. This effectively lowers the velocity of the material entering the main shock, therewith lowering the post-shock temperature \citep{Drury2009,Hughes0102, Vink2010}. Hence, $v_{\rm s}$ will not necessarily be the only variable to characterize $T_{\rm p}$. As we focus on $T_{\rm e}$-$T_{\rm p}$ equilibration, we plot $T_{\rm e}$ as function of the measured $T_{\rm p}$. This is more reliable as $v_{\rm s}$ is a quantity derived through Equation \ref{heat}, which may not be valid in the presence of cosmic-ray acceleration \citep{Helder2009}.

Fig. \ref{temperatures} shows that $T_{\rm e}\sim T_{\rm p}$ for the data point with the lowest proton temperatures. Surprisingly, the North $T_{\rm e}$ seems to be very high for the $T_{\rm p}$ based on the data from \cite{Ghavamian2007}. However, \cite{Long} measured approximately the same position and found a broader broad component (overplotted in green in Fig. \ref{temperatures}). The rightmost data points in Fig. \ref{temperatures} show  $T_{\rm e}<T_{\rm p} $, consistent with previous results, however, we do not find a relation of $T_{\rm e}/T_{\rm p} \propto 1/v_{\rm s}^2$, which would result in a horizontal line, showed as a dashed line (Fig. \ref{temperatures}).

 \section{Conclusions}
We investigated the proton-electron temperature equilibration behind several shock fronts of RCW~86. The low ionization parameter $n_{\rm e}t$ of this remnant ensures that the current $T_{\rm e}$ of the plasma is close that at the shock front. The electron temperatures were determined from X-ray spectra, obtained from XMM-Newton data and corrected for equilibration effects, resulting in Fig. \ref{temperatures}. We used both new and published data to determine proton temperatures. From this study, we can draw the following conclusions:

\begin{itemize}
\item[$-$] The FWHM of the broad H$\alpha$ line in the SE of RCW~86 are $1120\pm 40$ and $920\pm50$ \kms\ for the SE$_{\rm out}$ and SE$_{\rm in}$ spectrum respectively.
\item[$-$] $T_{\rm e}/T_{\rm p}\sim 1$ for the slow shocks in the E and for the N parameters of \cite{Long}. The N parameters of \cite{Ghavamian2007} and the SW form an exception to this. 
\item[$-$]  $T_{\rm e}/T_{\rm p}< 1$ for faster shocks. However, we do not find a constant electron temperature for faster shocks, as implied by the model of \cite{Ghavamian2007}. Additionally, our results show $T_{\rm e}/T_{\rm p} > m_{\rm e}/m_{\rm p}$.
\item[$-$] $T_{\rm e}/T_{\rm p}$ for the NE region, in which cosmic-ray acceleration appears to be efficient, is not well determined. However, we have found evidence for relatively high values of $T_{\rm e}$, suggesting that $T_{\rm e}/T_{\rm p}$ is close to one is this region. This may be attributed to efficient cosmic-ray acceleration, as this tends to lower the Mach number of the main shock. 

\end{itemize}

\section{Acknowledgements}
We thank Andrei Bykov and John Raymond for useful discussions on shock physics and the interpretation of optical and X-ray spectra. E.A.H. and J.V. are supported by the Vidi grant of J.V. from the Netherlands Organization for Scientific Research (NWO).

\bibliographystyle{apj}

\end{document}